\documentclass[a4paper,12pt]{article}

\usepackage{subfigure}
\usepackage{amssymb}
\usepackage{amsmath}
\usepackage{amsfonts}
\usepackage{graphicx}
\usepackage{psfrag}

\begin{document}

{\normalsize \hfill ITP-UU-10/07}\\
\vspace{-1.5cm}
{\normalsize \hfill SPIN-10/07}\\
${}$\\

\begin{center}
\vspace{24pt}
{ \Large \bf Causal Dynamical Triangulations\\ \vspace{10pt} 
and the\\ \vspace{18pt} 
Quest for Quantum Gravity}

\vspace{40pt}

{\sl J. Ambj\o rn}$\,^{a,c}$,
{\sl J. Jurkiewicz}$\,^{b}$
and {\sl R. Loll}$\,^{c}$

\vspace{24pt}
{\footnotesize

$^a$~The Niels Bohr Institute, Copenhagen University\\
Blegdamsvej 17, DK-2100 Copenhagen \O , Denmark.\\
{ email: ambjorn@nbi.dk}\\

\vspace{10pt}

$^b$~Institute of Physics, Jagellonian University,\\
Reymonta 4, PL 30-059 Krakow, Poland.\\
{ email: jurkiewicz@th.if.uj.edu.pl}\\

\vspace{10pt}

$^c$~Institute for Theoretical Physics, Utrecht University, \\
Leuvenlaan 4, NL-3584 CE Utrecht, The Netherlands.\\
{ email: r.loll@uu.nl}\\

\vspace{10pt}
}
\vspace{48pt}

\end{center}

\begin{center}
{\bf Abstract}
\end{center}

Quantum Gravity by Causal Dynamical Triangulation has over the last
few years emerged as a serious contender for a nonperturbative description 
of the theory. It is a nonperturbative implementation of the sum-over-histories,
which relies on few ingredients and initial assumptions, has few
free parameters and -- crucially -- is amenable to numerical simulations. 
It is the only approach to have demonstrated that a classical universe can
be generated dynamically from Planckian quantum fluctuations. 
At the same time, it allows for the explicit evaluation of 
expectation values of invariants 
characterizing the highly nonclassical, short-distance
behaviour of spacetime. As an added bonus, we have learned important
lessons on which aspects of spacetime need to be fixed a priori as part of
the background structure and which can be expected to emerge dynamically.

\newpage

\section{\large Quantum gravity - taking a conservative stance}

Many fundamental questions about the nature of space, time and gravitational
interactions are not answered by the classical theory
of general relativity, but lie in the realm of the still searched-for {\it theory of
quantum gravity}: What is the quantum theory underlying general relativity, and
what does it say about the quantum origins of space, time and our
universe? What is the microstructure of spacetime at the shortest scale usually
considered, the Planck scale $\ell_{\rm Pl}=10^{-35}m$, and what are the relevant
degrees of freedom determining the dynamics there? Are they the geometric
dynamical variables of the classical theory (or some short-scale version thereof), 
or do they also include the topology and/or dimensionality of spacetime, quantities
that classically are considered fixed? Can the dynamics of these microscopic 
degrees of freedom {\it explain} the observed large-scale structure of our own
universe, which resembles a de Sitter universe at late times? Do notions like 
``space", ``time" and ``causality" remain meaningful on short scales, or are they
merely macroscopically {\it emergent} from more fundamental, underlying Planck-scale principles?

Despite considerable efforts over the last several decades, 
it has so far proven difficult to come up with a 
consistent and quantitative theory of quantum gravity, which would be able to
address and answer such questions \cite{kiefer}. 
In the process, researchers in high-energy 
theory have been led to consider ever more radical possibilities in order to resolve 
this apparent impasse, from postulating the existence of extra structures 
unobservable at low energies to invoking ill-defined ensembles of multiverses 
and anthropic principles \cite{ellis}. 
A grand unified picture has quantum gravity inextricably
linked with the quantum dynamics of the three other known fundamental 
interactions, which requires a new unifying principle. Superstring theory is an
example of such a framework, which needs the existence of an as yet unseen 
symmetry (supersymmetry) and ingredients (strings, branes, fundamental
scalar fields). Loop quantum gravity, a non-unified approach, postulates the
existence of certain fundamental quantum variables of Wilson loop type. Even
more daring souls contemplate -- inspired by quantum-gravitational problems --
the abandonment of locality \cite{giddings} or substituting quantum mechanics
by a more fundamental, deterministic theory \cite{gerard}. 
   
In view of the fact that none of these attempts has as yet thrown much light on the
questions raised above, and that we have currently neither direct tests of 
quantum gravity 
nor experimental facts to guide our theory-building, a more
conservative approach may be called for. What we will sketch in the
following is an alternative route to quantum gravity, which relies on
nothing but standard principles from quantum field theory, and on ingredients and
symmetries already contained in general relativity. Its main premise is that
{\it the framework of standard quantum field theory is sufficient to construct and
understand quantum gravity as a fundamental theory,} if {\it the dynamical,
causal and nonperturbative properties of spacetime are taken into account properly.}


Significant support for this thesis comes from a new candidate theory,
{\it Quantum Gravity from Causal Dynamical Triangulation (CDT)}, whose 
main ideas and results will be described below. CDT quantum gravity is a 
nonperturbative implementation of the gravitational path integral, and has
already passed a number of nontrivial tests with regard to producing the
correct classical limit. Its key underlying idea was conceived more than
ten years ago \cite{al}, in an effort to combine the insights of geometry-based
nonperturbative canonical quantum gravity with the powerful calculational and
numerical methods available in covariant approaches. After several years
of modelling and testing both the idea and its implementation in 
spacetime dimensions two and three, where they give rise to nontrivial 
dynamical systems of ``quantum
geometry" \cite{loeu,3dcdt}, the first results for the physically relevant case of
four dimensions were published in 2004 \cite{ajl-prl, ajl-rec}.  

Let us also mention that an independent approach to the 
quantization of gravity, much in the spirit 
of our main premise\footnote{although the role of ``causality", which 
enters crucially in CDT quantum gravity, remains unclear in this approach} and 
based on the 30-year-old idea of ``asymptotic
safety" has been developing over roughly the same time period \cite{niereu,
niedermaier}. It shares some features (covariance, amenability to numerical
computation) as well as some results (on the spectral dimension) with CDT
quantum gravity, and may ultimately turn out to be related.

\section{\large What CDT quantum gravity is about}

Quantum gravity theory based on causal dynamical triangulations is an
explicit, nonperturbative and background-independent realization of the formal 
{\it gravitational 
path integral} (a.k.a. the ``sum over histories") on a differential manifold $M$,
\begin{equation}
Z(G_N,\Lambda)=
\int\limits_{ {\cal G}({\rm M})=\frac{\rm Lor(M)}{\rm Diff(M)}}
{\cal D}g_{\mu\nu}\ {\rm e}^{iS^{\rm EH}[g_{\mu\nu}]}, \;\;\;\;
S^{\rm EH}=\int d^4x \sqrt{\det g} (\frac{1}{G_N} R-2 \Lambda),
\label{gravint}
\end{equation}
where $S^{\rm EH}$ denotes the four-dimensional Einstein-Hilbert action,
$G_N$ is the gravitational or Newton's constant and 
$\Lambda$ the cosmological constant, and the path integral is to be taken over 
all spacetimes (metrics $g_{\mu\nu}$ modulo diffeomorphisms), with specified 
boundary conditions.  
The method of causal dynamical triangulation 
turns (\ref{gravint}) into a well-defined finite and regularized 
expression, which can be evaluated and whose continuum limit (removal
of the regulator) can be studied systematically \cite{ajl1}.

\begin{figure}[t]
\centerline{\scalebox{0.5}{\rotatebox{0}{\includegraphics{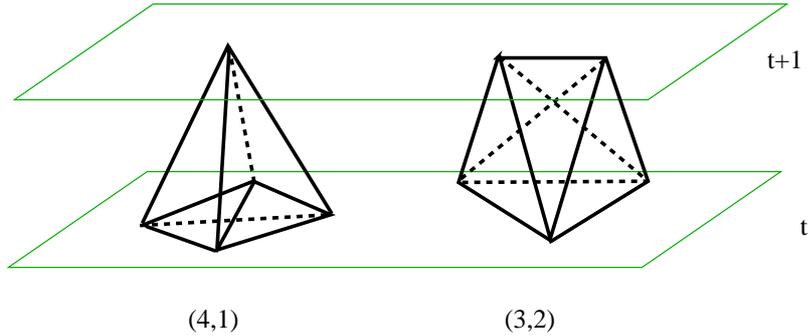}}}}
\caption{The two fundamental building blocks of CDT are four-simplices with
flat, Minkowskian interior. They are spanned by spacelike edges, 
which lie entirely within spatial slices of constant
time $t$, and timelike edges, which interpolate between adjacent slices of
integer time. A building block of type $(m,n)$ has $m$ of its vertices in
slice $t$, and $n$ in slice $t+1$.
}
\label{buildingblocks}
\end{figure}

One proceeds in analogy with
the path-integral quantization \`a la Feynman and Hibbs of the nonrelativistic 
particle. This is
defined as the continuum limit of a regularized sum over paths,
where the contributing `virtual' paths are taken from an ensemble of piecewise
straight paths, with the length $a$ of the individual segments going to zero in the limit.
The corresponding CDT prescription in higher dimensions is to represent the 
space $\cal G$ of all Lorentzian spacetimes 
in terms of a set of triangulated, 
piecewise flat manifolds\footnote{Unlike in the particle case,
there is no embedding space; all geometric spacetime data are defined 
intrinsically, just like in the classical theory.}, as originally introduced in the classical
theory as ``general relativity without coordinates" \cite{regge}. For our 
purposes, the simplicial approximation ${\cal G}_{a,N}$ of $\cal G$ contains
all simplicial manifolds $T$ obtained from gluing together at most 
$N$ four-dimensional,
triangular building blocks of typical edge length $a$, with $a$ again playing
the role of an ultraviolet (UV) cut-off (see Fig.\ \ref{buildingblocks}). 
The explicit form of the regularized gravitational path integral in CDT is   
\begin{equation}
\label{discretesum}
Z^{\rm CDT}_{a,N}= \sum_{{\rm triangulated} \atop 
{\rm spacetimes} \, T\in{\cal G}_{a,N}}\frac{1}{C_T}{\rm e}^{i S^{\rm Regge}[T]},
\end{equation}
where $S^{\rm Regge}$ is the Regge version of the Einstein-Hilbert action 
associated with the simplicial spacetime $T$, and  
$C_T$ denotes the order of its automorphism group.
The discrete volume $N$ acts as a volume cutoff. 
We still need to consider a suitable continuum or scaling limit 
\begin{equation}
\label{discretelim}
Z^{\rm CDT}:=\lim_{N\rightarrow\infty\atop a\rightarrow 0} Z^{\rm CDT}_{a,N}
\end{equation}
of (\ref{discretesum}), while renormalizing the original bare coupling constants
of the model, in order to arrive (if all goes well) at a theory of quantum gravity.
The two limits in (\ref{discretelim}) are
usually tied together by keeping a physical four-volume, defined as $V_4:=a^4 N$ fixed. 
In the limiting process $a$ is taken to zero, $a\rightarrow 0$, and the individual discrete building blocks 
are then literally ``shrunk away". 

Let us summarize the key features of the construction scheme thus introduced.
Unlike what is possible in the continuum theory, the path integral 
(\ref{discretesum}) is defined directly on the physical configuration space of
{\it geometries}. It is nonperturbative in the sense of including 
geometries which are ``far away" from any classical solutions, and it is
background-independent in the sense of performing the sum ``democratically",
without distinguishing any given geometry (say, as a preferred background). 
Of course, these nice properties of the regularized path integral are only 
useful because \textit{we are able to evaluate $Z^{\rm CDT}$ quantitatively},
with an essential role being played by Monte Carlo simulations. These, together
with the associated finite-size scaling techniques \cite{newmanbarkema}, have
enabled us to extract information about the nonperturbative, 
strongly coupled quantum
dynamics of the system which is currently not accessible by analytical
methods, neither in this nor any other approach to quantum gravity. 
It is reminiscent of the role played by lattice simulations in pinning down
the nonperturbative behaviour of QCD (although this is a theory we 
already know {\it much} more about than quantum gravity).

\section{\large What CDT quantum gravity is not about}

Although causal dynamical triangulation is sometimes called a discrete
approach, this is potentially misleading. First, one can of course think of the simplicial
building blocks as discrete objects, but they are assembled into spacetimes
that are perfectly continuous and not discrete. The space of geometries 
{\it is} discretized in the sense that both four-volume and curvature contribute 
in discrete ``bits" to the total action. However, this is only a feature of the 
chosen regularization, and has no physical significance as such. 
As explained in the previous section, the characteristic
edge length $a$ plays the role of an intermediate regulator and UV cut-off for
the geometry. In the continuum limit, $a$ is to be taken to zero strictly. 
In practice, what will usually suffice is to
choose $a$ significantly smaller than the scale at which one is trying to extract physical
results, hence $a\ll \ell_{\rm Pl}$ if we want to establish Planck scale dynamics.

Adherents of the idea of fundamental discreteness might be tempted to
identify the edge length $a$ with a fundamental, shortest length scale, typically,
the Planck length. However, this would be an ad hoc prescription 
which is in no way required by the construction. Besides, it has the unpleasant
feature that physics at the Planck scale will then depend explicitly on the details
of the chosen regularization. For example, choosing squares instead of triangles,
or choosing a different discrete realization of the Einstein action will in general lead to
different Planckian dynamics, thus introducing an infinite ambiguity {\it at that scale}.
It is not good enough if all these different theories
produce identical classical physics on large
scales, because in quantum gravity one is of course interested in finding
a (hopefully unique) description of physics at the Planck scale.

Instead of putting it in by hand, the issue of fundamental discreteness in
quantum gravity needs to be addressed {\it dynamically}. Is such a scale
generated by the dynamics of the theory? 
Although there are numerous claims that Planck-scale discreteness is almost
``self-evident" (often, to render one's favourite calculation of black hole 
entropy finite), there is at this stage no concrete evidence for such a discreteness in full, 
four-dimensional quantum gravity\footnote{The derivation of discrete 
aspects of the spectrum of the area and
volume operators in loop quantum gravity \cite{discreteness,Lollvol} disregards 
dynamics (in the form of the Hamiltonian constraint), quite apart from
the fact that one can argue that discreteness has been put in ``by hand"
by choosing a quantum representation where one-dimensional Wilson loops are well-defined
operators.}. We have up to now not 
seen any indication of it, but it is conceivable that there exist 
nonperturbative quantum operators in CDT quantum gravity
which measure lengths (or higher-dimensional volumes) and have a discrete
spectrum as $a\rightarrow 0$, thus indicating fundamental discreteness. 
Even if such a discreteness were found, whether or not the currently unknown ``fundamental
excitations of quantum gravity" are discrete or not may be yet another issue.
It is not even particularly clear what one means by such a statement and 
whether
it can be turned into an operationally well-defined question in the
nonperturbative theory, and not one which is merely a feature of a particular
representation of the quantum theory.

\vspace{0.3cm}

As we will see in more
detail below, CDT is -- as far as we are aware -- the only 
nonperturbative approach to quantum gravity 
which has been able to dynamically generate its own, physically realistic
background from nothing but quantum fluctuations. More than that, because
of the minimalist set-up and the methodology used (quantum field theory
and critical phenomena), the results obtained are robust in the sense of
being largely independent of the details of the chosen regularization procedure 
and containing few free parameters. This is therefore also one of the perhaps 
rare instances of a candidate theory of quantum gravity which can potentially
be falsified. In 
fact, the Euclidean version of the theory extensively studied in the 1990s 
has already been falsified because it does not lead to the correct classical
limit \cite{bialas,debakker}. CDT quantum gravity improves on this previous
attempt by building a causal structure right into the fabric of the model.    
Our investigations of both the quantum properties and the classical limit
of this candidate theory are at this stage not sufficiently complete to provide 
conclusive evidence that we have found {\it the} correct theory of quantum gravity,
but results until now have been unprecedented and most encouraging, and
have thrown up a number of nonperturbative surprises.

\section{\large CDT key achievements I - Demonstrating the need for causality}

We will confine ourselves to highlighting
some of the most important results and new insights obtained in CDT quantum
gravity, without entering into any of the technical details. The reader is
referred to the literature cited in the text, as well as to the various
overview articles available on the subject \cite{review} for more information.

The crucial lesson learned for nonperturbative
gravitational path integrals from CDT quantum gravity is that the ad hoc
prescription of integrating over curved Euclidean {\it spaces} of metric signature 
(++++) instead of the physically correct curved Lorentzian space{\it times}
of metric signature ($-$+++) generally leads to inequivalent and (in $d\! =\! 4$)
incorrect results. ``Euclidean quantum gravity" of this kind, as advocated
by S. Hawking and collaborators \cite{eqg}, adopts this version of
doing the path integral mainly for the technical reason to be able to 
use real weights $\exp (-S^{\rm eu})$ instead of the complex amplitudes
$\exp (i S^{\rm lor})$ in its evaluation. The same prescription is used routinely
in perturbative quantum field theory on flat Minkowski space, but in that
case one can rely on the existence of a well-defined Wick rotation to 
relate correlation functions in either signature. This is {\it not} available
in the context of continuum gravity beyond perturbation theory on a
Minkowski background, but one may still hope that by {\it starting
out} in Euclidean signature and quantizing this (wrong) theory, an inverse 
Wick rotation
would then ``suggest itself" to translate back the final result into physical,
Lorentzian signature. Alas, this has never happened, because -- we would
contend -- no one has
been able to make much sense of nonperturbative 
Euclidean quantum gravity in the
first place, even in a reduced, cosmological context\footnote{a discussion of
the kind of problems that arise can be found in \cite{louko}}.

\begin{figure}[t]
\centering
\subfigure{\includegraphics[scale=0.3]{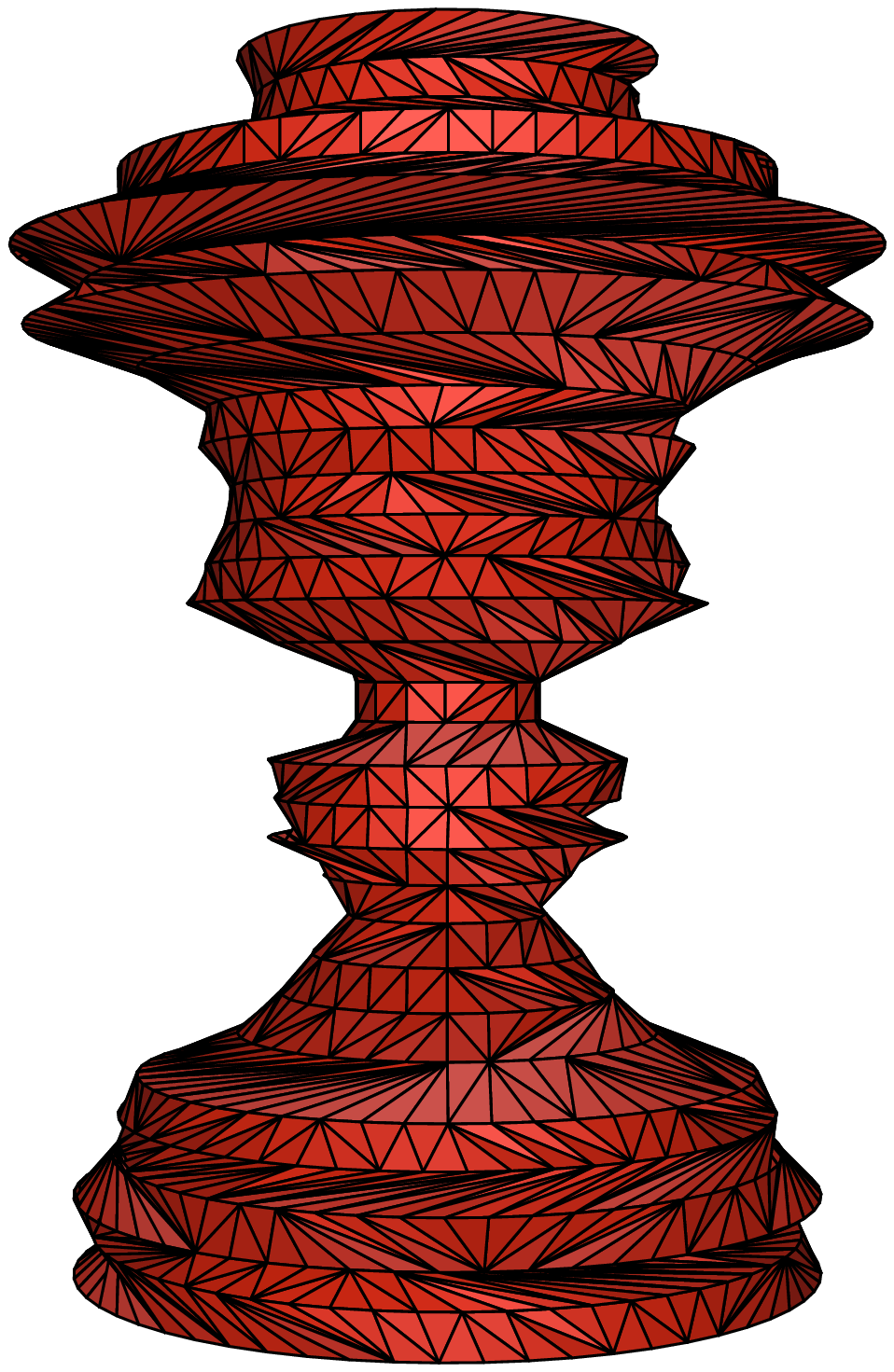}}%
\subfigure{\includegraphics[scale=0.45]{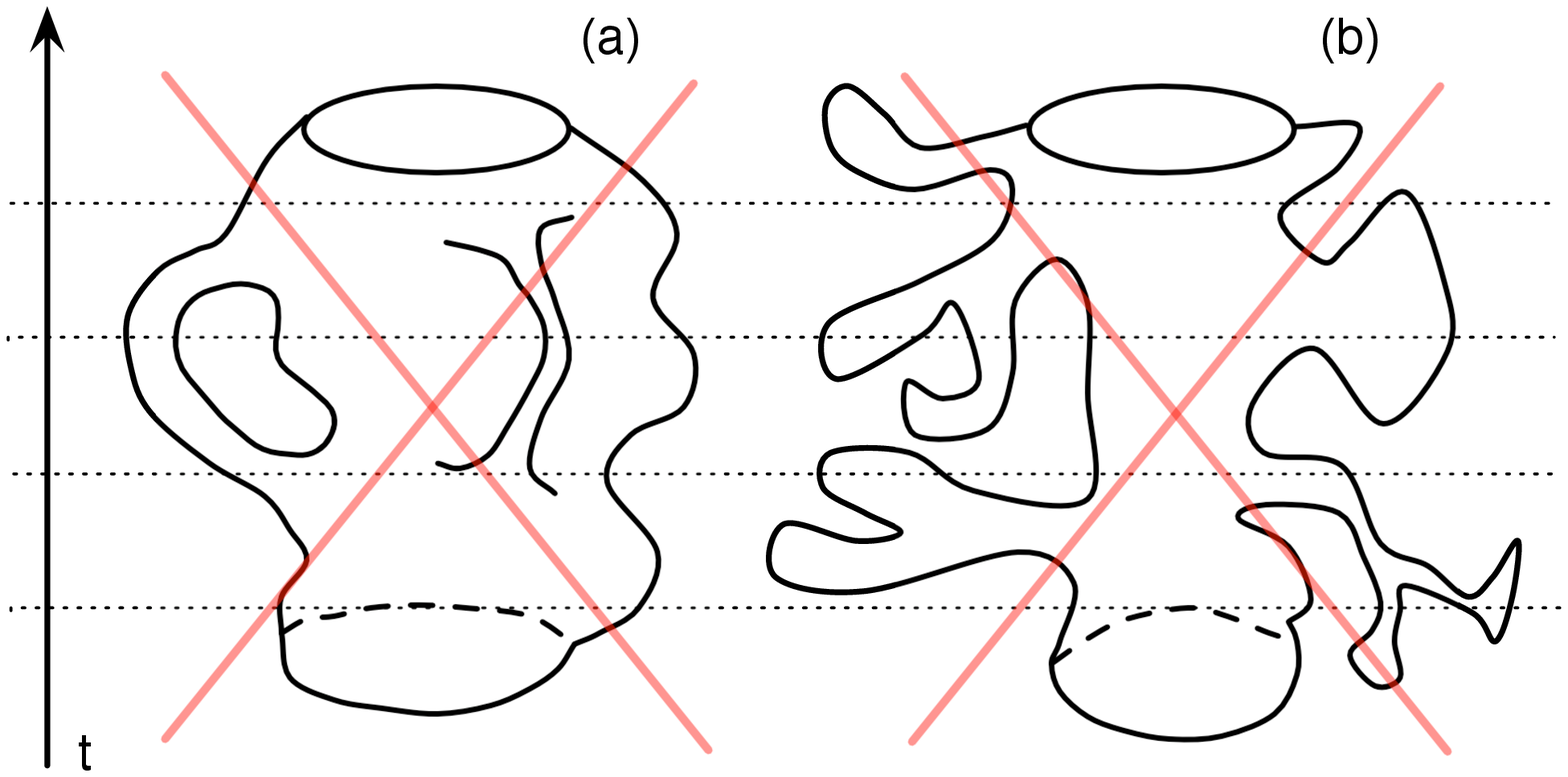}}%
\caption{Typical history contributing to the loop-loop correlator in the
2d Lorentzian CDT path integral (left), time $t$ is pointing up. The essential
difference with the corresponding Euclidean amplitude is that the 
(one-dimensional) spatial slices, although quantum-fluctuating,
are not allowed to change topology as a function of time $t$, thus avoiding
causality-violating branching and merging points. This excludes spaces with 
wormholes
(right picture, a) and those with `baby universes' branching out in the time 
direction (right picture, b).}
\label{lor2d}
\end{figure}
CDT quantum gravity has provided the first explicit example of a nonperturbative
gravitational path integral (in a toy model of two-dimensional gravity) which is exactly
soluble and leads to distinct and inequivalent results depending on whether
the sum over histories is taken over Euclidean spaces or Lorentzian spacetimes
(or, more precisely, Euclidean spaces which are obtained by Wick rotation --
which {\it does} exist for the class of simplicial spacetimes under
consideration -- from Lorentzian spacetimes). The Lorentzian path integral
was first solved in \cite{al}, and a quantity one can compute
and compare with the Euclidean version found in \cite{multiloop} is the 
cylinder amplitude (Fig.\ \ref{lor2d}).
In the Lorentzian CDT case, only those histories are summed over which 
possess a global time slicing {\it with respect to which no spatial topology
changes are allowed to occur}. After Wick rotation, this set constitutes a
strict subset of all Euclidean (triangulated) spaces. In the latter there is no natural
notion of `time' or `causality' and branching geometries are thus always present.

In the two-dimensional setting this means that one has identified a new
class of anisotropic statistical mechanical models of fluctuating geometry. 
Several intriguing results that have been found from numerical simulations of 
matter-coupled versions
of the model (which have so far resisted analytical solution) indicate that
their geometric disorder is less severe than that of their Euclidean
counterparts, and in particular that they seem to lead to critical matter
exponents identical to those of the corresponding matter model on
fixed, flat lattices \cite{spin,potts}. 

Another new direction in which the two-dimensional CDT model has been
generalized is a controlled relaxation of the ban on branching points, while
adhering to a global notion of proper time \cite{topo,tame}. This has
culminated recently in the formulation of a fully-fledged CDT string
field theory in zero target space dimensions \cite{cdtsft}. The matrix model
formulation of the theory makes it possible to perform the sum over
two-dimensional topologies explicitly \cite{sumtopo}.
These developments are described in more detail elsewhere in this volume
\cite{stoch}.

\begin{figure}[t]
\centering
\subfigure{\includegraphics[scale=0.4]{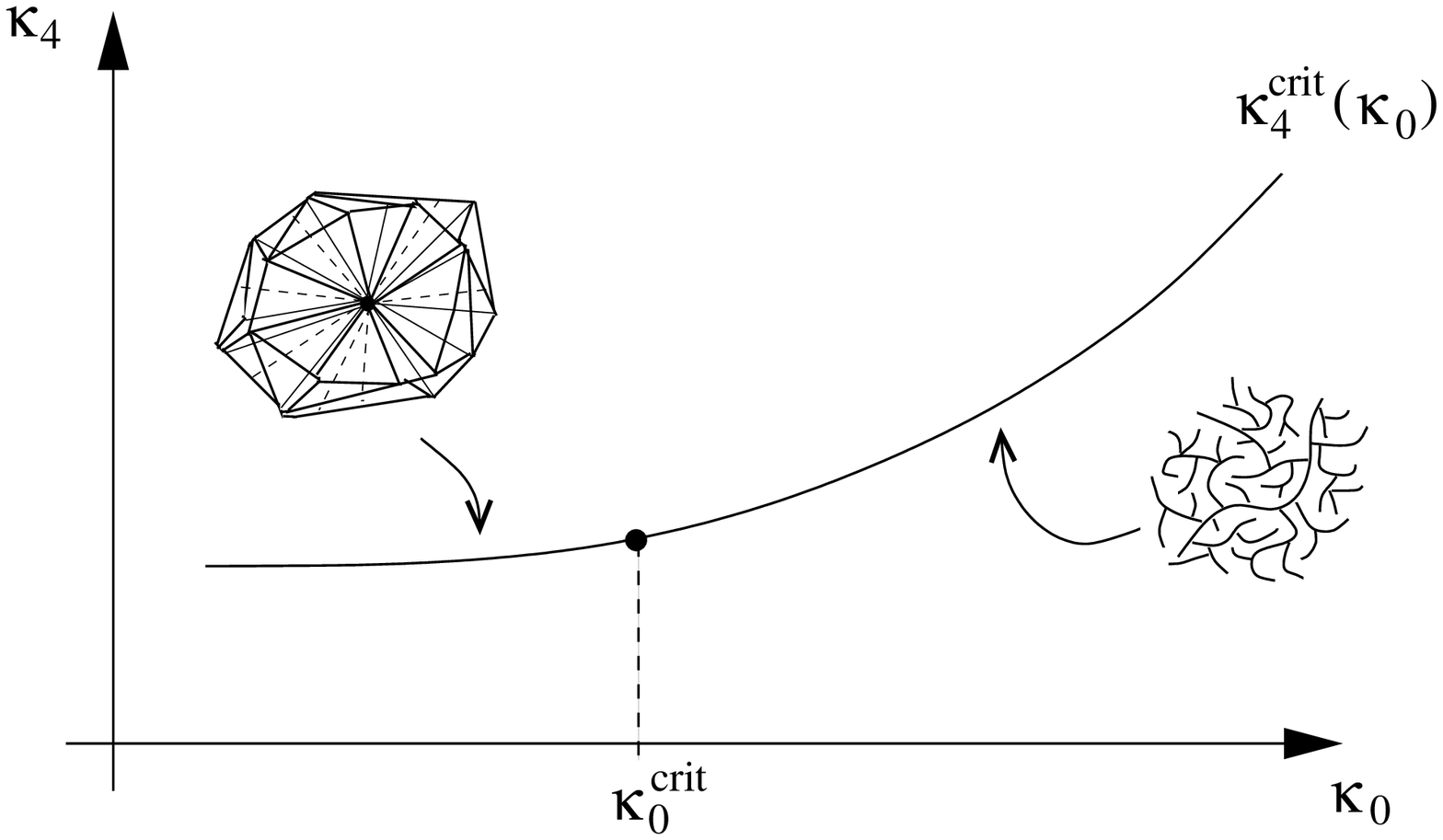}}%
\hspace{1cm}
\subfigure{\includegraphics[scale=0.35]{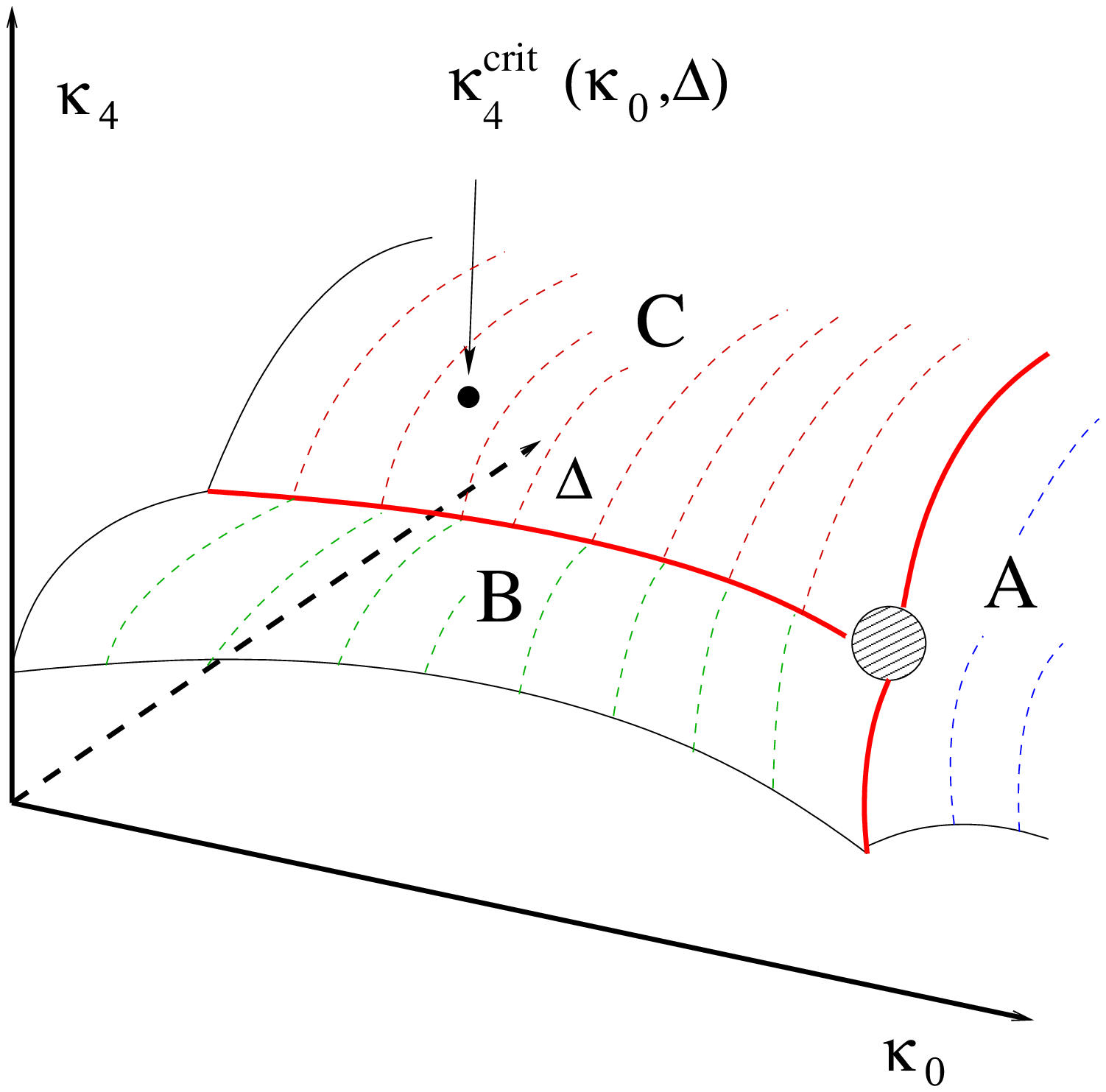}}%
\caption{The phase diagrams of Euclidean (left) and Lorentzian (right) quantum
gravity from dynamical triangulations, with $\kappa_0$ and $\kappa_4$ denoting
the bare inverse Newton's constant and (up to an additive shift) the bare
cosmological constant. After fine-tuning to the respective subspace where 
the cosmological
constant is critical (tantamount to performing the infinite-volume limit), there are
(i) two phases in EDT: the crumpled phase $\kappa_0<\kappa_0^{\rm crit}$
with infinite Hausdorff dimension and the branched-polymer phase 
$\kappa_0>\kappa_0^{\rm crit}$ with Hausdorff dimension 2, none of them with
a good classical limit, (ii) three phases in CDT: A and B (the Lorentzian 
analogues of the branched-polymer and crumpled phases), and a {\it new}
phase C, where an extended, four-dimensional universe emerges. The parameter
$\Delta$ in CDT parametrizes a finite relative scaling between space- and
time-like distances which is naturally present in the Lorentzian case.}
\label{phases}
\end{figure}
Returning to the implementation of strict causality on path integral histories,
a key finding of CDT quantum gravity is that a result similar to that found in
two dimensions also holds also in dimension four.
The geometric degeneracy of the phases (in the sense of 
statistical systems) found in Euclidean dynamical triangulations \cite{bialas1,bialas}, 
and the
resulting absence of a good classical limit can in part be traced to the
`baby universes' present in the Euclidean approach also in four dimensions. 
As demonstrated
by the results in \cite{ajl-prl,ajl-rec}, the requirement of microcausa\-lity (absence
of causality-violating points) of the individual path integral histories leads to
a qualitatively new phase structure, containing a phase where the universe on 
large scales is extended and four-dimensional (Fig.\ \ref{phases}), as required
by classical general relativity.
Apart from the nice result that the problems of the Euclidean approach are
cured by this prescription, this reveals an intriguing relation between 
the microstructure of spacetime (micro-causality = suppression of baby
universes in the time direction at sub-Planckian and bigger scales) and its emergent
macrostructure. Referring to the questions raised at the beginning of Sec.\ 1,
the more general lessons learned from this are that (i) ``causality" is not emergent,
but needs to be put in by hand on each spacetime history, and (ii) similarly, ``time"
is not emergent. It is put into CDT by choosing a preferred (proper-)time
slicing at the regularized level, but this turns out to be only a necessary
condition to have a notion of time (as part of an extended
universe) present in the continuum limit, at least on large scales. 
It is not sufficient, because in other phases of the CDT model 
(Fig.\ \ref{phases}) the spatial universe apparently does not persist at all (B)
or only intermittently (A), see also reference \cite{ajl-rec}.

\section {\large CDT key achievements II - The emergence of spacetime as we know it}

This brings us straight to the nature of the extended spacetime found in
phase C of CDT quantum gravity. What is it, and how do we know? We cannot
just `look at' the quantum superposition of geometries, which individually of
course get wilder and spikier as the continuum limit $a\rightarrow 0$ is
approached, just like the nowhere differentiable paths of the path integral
of the nonrelativistic particle \cite{reedsimon}. We need to define and measure
geometric {\it quantum observables}, evaluate their expectation values on the
ensemble of geometries and
draw conclusions about the behaviour of the ``quantum geometry" generated
by the computer simulations (that is, the ground state of minimal Euclidean
action). Rather strikingly, inside phase C the many microscopic building blocks
superposed in the nonperturbative path integral `arrange themselves' into an
extended quantum spacetime whose macroscopic shape is that of the well-known
{\it de Sitter universe} \cite{desitter,desitter1}. This amounts to a highly nontrivial
test of the classical limit, which is notoriously difficult to achieve 
in models of nonperturbative quantum gravity.
The precise dynamical mechanism by which this
happens is unknown, however, it is clear that ``entropy" (in other words,
the measure of the path integral, or the number of times a given weight factor
$\exp(-S)$ is realized) plays a crucial role in producing the outcome. This is
reminiscent of phenomena in condensed matter physics, where systems of 
large numbers of microscopic, interacting constituents exhibit macroscopic,
``emergent" behaviour which is difficult to derive from the microscopic laws
of motion. This makes it appropriate to think of CDT's de Sitter space as a {\it self-organizing 
quantum universe} \cite{SO}. 
  
\begin{figure}[t]
\centering
\vspace*{10pt}
\includegraphics[width=9cm]{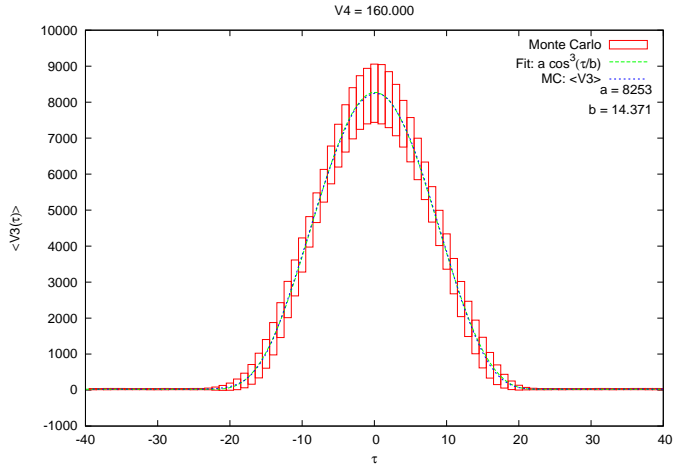}
\vspace*{10pt}
\caption{The shape $\langle V_3(\tau)\rangle$ of the CDT quantum universe, fitted 
to that of Euclidean de Sitter space (the ``round four-sphere") with rescaled proper time, $\langle V_3(\tau)\rangle=a \cos^3(\tau/b)$. Measurements taken for a universe of 
four-volume $V_4=160.000$ and time extension $T=80$. The fit of the Monte Carlo 
data to the theoretical curve for the given values of $a$ and $b$ is impressive. 
The vertical boxes quantify the typical scale of quantum fluctuations around 
$\langle V_3(\tau)\rangle$.}
\label{spherefit}
\end{figure}
The manner in which we have identified (Euclidean) de Sitter space from the
computer data is by looking at the expectation value of the
volume profile $V_3(t)$, that is, the
size of the spatial three-volume as function of proper time $t$. For a
classical Lorentzian de Sitter space this is given by
\begin{equation}
\label{profile}
V_3(t)= 2 \pi^2 (c \cosh \frac{t}{c})^3, \;\; c=const. >0,
\end{equation}
which for $t>0$ gives rise to the familiar, exponentially expanding universe,
thought to give an accurate description of our own universe at late times,
when matter can be neglected compared with the repulsive
force due to the positive cosmological constant. Because the CDT simulations
for technical reasons have to be performed in the Euclidean regime, we
must compare the expectation value of the shape with those of the
analytically continued expression of (\ref{profile}), with respect to the
Euclidean time $\tau:=-i t$. After normalizing the overall four-volume and
adjusting computer proper time by a constant to match continuum proper
time, the averaged volume profile is depicted in Fig.\ \ref{spherefit}.

A few more things are noteworthy about this result. Firstly, despite the
fact that the CDT construction deliberately breaks the isotropy between
space and time, at least on large scales the full isotropy is restored by the
ground state of the theory for precisely one choice of identifying 
proper time, that is, of fixing a relative scale between time and spatial distances 
in the continuum. Secondly, the computer simulations by
necessity have to be performed for finite, compact spacetimes, which also
means that a specific choice has to be made for the spacetime topology. 
For simplicity, to avoid having to specify boundary conditions, it is usually
chosen to be $S^1\times S^3$, with time compactified\footnote{the period
is chosen much larger than the time extension of the universe and does
not influence the result} and spatial slices which are topological three-spheres.
What is reassuring is the fact that the bias this could in principle have introduced 
is ``corrected" by the system, which clearly is driven dynamically to the
topology of a four-sphere (as close to it as allowed by the kinematical constraint
imposed on the three-volume, which is not allowed to vanish at any time).
Lastly, we have also analyzed the quantum fluctuations around the
de Sitter background - they match to good accuracy a continuum
saddlepoint calculation in minisuperspace \cite{desitter1}, which is
one more indication that we are indeed on the right track. 

\section {\large CDT key achievements III - a window on Planckian dynamics}

Having discussed some of the evidence for obtaining the correct classical
limit in CDT quantum gravity, let us turn to the {\it new} physics we are
after, namely, what happens to gravity and the structure of spacetime at
or near the Planck scale. We will describe one way of probing the short-scale 
structure, by setting up a {\it diffusion process} on the ensemble of spacetimes, 
and studying associated observables. The speed by which an initially localized 
diffusion process spreads into an ambient space is sensitive to the dimension
of the space. Conversely, given a space $M$ of unknown properties, it can be
assigned a so-called {\it spectral dimension} $D_S$ by studying the
leading-order behaviour of the average return probability ${\cal R}_V(\sigma)$
(of random diffusion
paths on $M$ starting and ending at the same point $x$) as a function of the
external diffusion time $\sigma$,
\begin{equation}
{\cal R}_V(\sigma):=\frac{1}{V(M)}\int_M d^dx\ P(x,x;\sigma)\propto 
\frac{1}{\sigma^{D_S/2}},\;\;\;\;\; \sigma \leq V^{2/D_S},
\end{equation}   
where $V(M)$ is the volume of $M$, and $P(x_0,x;\sigma)$ the solution to
the heat equation on $M$,
\begin{equation}
\partial_\sigma P(x_0,x,\sigma)=\nabla_x^2 P(x_0,x,\sigma).
\label{diffeq}
\end{equation}
\begin{figure}[t]
\centerline{\scalebox{0.8}{\rotatebox{0}{\includegraphics{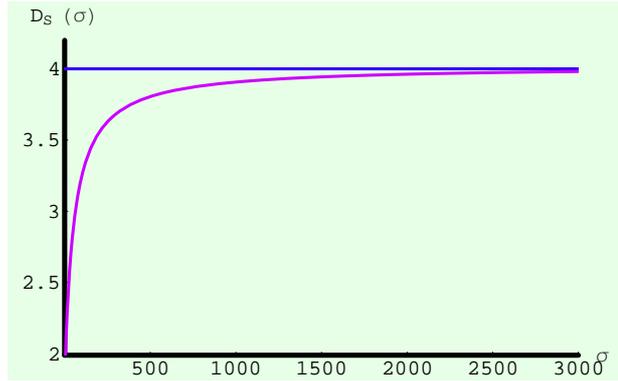}}}}
\caption{The spectral dimension $D_S(\sigma)$ of the CDT-generated 
quantum universe (lower curve, error bars not included), contrasted
with the corresponding curve for a classical spacetime, simply given by the
constant function $D_S(\sigma)=4$. We assume $\sigma \ll V^{2/D_S}$, so
no finite-volume effects are present.
}
\label{specdimnew}
\end{figure}
Diffusion processes can be defined on very
general spaces, for example, on fractals, which are partially characterized
by their spectral dimension (usually not an integer, see \cite{fractals}). Relevant
for the application to quantum gravity is that the expectation value
$\langle {\cal R}_V(\sigma)\rangle$ can be measured on the ensemble of
CDT geometries, giving us the spectral dimension of the dynamically
generated quantum universe, with the result that $D_S(\sigma)$
depends nontrivially on the diffusion time $\sigma$ \cite{spectral}! 
Since the linear scale probed in the diffusion is on the order of 
that of a random walker, $\sqrt{\sigma}$, short diffusion times probe
the short-scale structure of geometry, and long ones its large-scale
structure. 
The measurements from CDT quantum gravity, extrapolated
to all values of $\sigma$, lead to the lower curve in Fig.\ \ref{specdimnew}, with
asymptotic values $D_S(0)=1.82\pm 0.25$, signalling highly 
nonclassical behaviour
near the Planck scale, and $D_S(\infty)=4.02\pm 0.1$, which is compatible
with the expected classical behaviour. Previous Euclidean models never showed
such a scale-dependence, reflecting their lack of an interesting geometric
structure as a function
of scale. For CDT in three space-time dimensions, there is evidence for an
analogous scale dependence \cite{3dspec}.

This somewhat unexpected result found in nonperturbative CDT
quantum gravity has brought into focus the role of ``dynamical dimensions" as
diffeomor\-phism-invariant indicators of nonclassicality at the Planck 
scale\footnote{Other 
notions of dimensionality are the Hausdorff and the fractal dimension.}. 
Interestingly, a similar
dimensional reduction from four to two near the Planck scale
has since been found in disparate approaches to quantum gravity,
most prominently, a nonperturbative renormalization group flow analysis 
\cite{laureu}, and so-called Lifshitz gravity \cite{Horava}. 
The coincidence is certainly intriguing and could mean that 
at a more fundamental level the approaches 
have more in common than we currently understand ({\it and} capture
a true aspect of nonperturbative quantum gravity). Reproducing dimensional 
reduction could then even become a ``test" of quantum gravity, similar to how
the derivation of the black hole entropy formula $S_{\rm BH}=A/4$ is 
often viewed, with the difference that the latter is usually associated with
a semiclassical context, whereas the former is thought to characterize
the behaviour of the theory in the deep UV.

Further evidence of nonclassicality
on short scales in CDT comes from measurements of geometric structures in
spatial slices $\tau=const$, including a measurement of 
their Hausdorff and spectral dimensions \cite{ajl-rec}.

\section{\large Open issues and outlook}

As we have summarized above, significant strides have been made in the
causal dynamical triangulations quantization program in 
demonstrating its compatibility with classical general relativity on large scales,
and at the same time exploring its true quantum properties on small scales.
A number of important issues are the subject of ongoing and future research. 
Firstly, as explained in more detail in \cite{desitter1}, one would like to
tune the bare parameters of the CDT simulations so as to obtain a better 
length resolution and get even closer to and, if possible, below the Planck scale.
(Current simulations operate with quantum universes of the order of 10-20
Planck lengths across.) This would enable us to look for more direct
evidence of the existence or otherwise of the nontrivial UV fixed point seen 
in truncated renormalization group flows \cite{laureu1}. 

An important challenge is
to reproduce further aspects of the classical limit correctly, one of which is the
derivation of Newton's law ``from scratch" in the nonperturbative theory. As
a possibly first step towards this goal, some physical consequences of the
presence of an isolated point mass in CDT's quantum de Sitter universe
have been analyzed in \cite{pointmass}. Another natural area of application
is the early universe, with or without the addition of a scalar field (``inflaton"),
to check and discriminate between the often ambiguous and 
contradictory claims of quantum cosmological models in a context where
{\it all} fluctuations of the geometry are present, not just the overall scale
factor. A concrete example of what one might be able to do is nailing down factor-ordering
ambiguities in the cosmological path integral \cite{factor}.  

\vspace{0.3cm}

Coming back to some of the questions we raised at the outset of this article,
the preliminary conclusion about the nature of
quantum spacetime is that it is 
nothing like a four-dimensional classical manifold on
short scales. In addition to its anomalous spectral dimension, its 
na\"ive Regge curvature diverges, indicating a singular behaviour
reminiscent of (but surely worse than) that of the particle paths constituting the
support of the Wiener measure. However, it apparently is {\it not} literally a 
``spacetime foam", if by that one means some bubbling, topology-changing
entity: one of the main findings of dynamically triangulated models of
nonperturbative quantum gravity is that allowing for local topology changes and
making them part of the dynamics renders the quantum superposition
inherently unstable and is incompatible with a good classical limit. 
Even if local topo\-lo\-gy change is not part of quantum-gravitational dynamics, 
we saw that global topology, as well as short-scale dimensionality are determined 
dynamically and do not necessarily coincide with the (somewhat arbitrary)
choices made for them as part of the regularized formulation. What makes 
these perhaps surprising findings possible is the fact that CDT quantum gravity
allows for large curvature fluctuations on short scales, and that the construction
of the final theory involves a nontrivial limiting process, which the computer
simulations are able to approximate.

In summary, if there is indeed a unique, interacting quantum field theory of
spacetime geometry in four dimensions, which does not contain any exotic
ingredients, and has general relativity as its classical limit, the CDT approach
has a good chance of finding it. It relies only on a minimal set of ingredients and
priors: the quantum superposition principle, locality, (micro-)causality, a notion
of (proper) time and standard tools from quantum field 
theory otherwise\footnote{This puts it about on a par with the renormalization group
approach of \cite{niereu,niedermaier}, makes fewer assumptions than
loop quantum gravity \cite{lqg}, but is not quite as minimalistic as the 
causal set approach \cite{cs}.}, has few free parameters 
(essentially the couplings of the
phase diagram of Fig.\ \ref{phases}), and by virtue of its construction through
a scaling limit can rely on a considerable degree of universality in the sense
of critical system theory. Although many issues remain to be tackled and
understood, the interesting new results and insights CDT has produced to date
make for a pretty good start.

\vspace{.3cm}

\noindent{\bf Acknowledgments.} All authors gratefully acknowledge
support by ENRAGE (European Network on
Random Geometry), a Marie Curie Research Training Network, 
contract MRTN-CT-2004-005616. In addition, 
JJ was supported by COCOS (Correlations in Complex Systems), 
a Marie Curie Transfer
of Know\-ledge Project, contract MTKD-CT-2004-517186, and
RL by the Netherlands
Organisation for Scientific Research (NWO) under their VICI
program.

\end{document}